# NONLINEAR PLANNING MODEL

## With a Gaussian Criterion of Optimization

### (Gaussian Programming Model)

**Mikhail Luboschinsky**

*We propose an Economic - Probabilistic analogy: the category of cost is analogous to the category of Probability. The proposed analogy permits construction of an informal theory of nonlinear non-convex Gaussian Utility and Cost, which describes the real economic processes more adequately than a theory based on a linear and convex models.*

*Based on the proposed analogy, we build a nonlinear non-convex planning model with a Gaussian optimality criterion – Gaussian Programming Model.*

*We also describe a corresponding model of Generalized Piecewise-Linear Programming that can be used to approximate a Gaussian Programming model, and vice versa.*

*Proposed constructions are illustrated on a numerical example.*

========================================

# НЕЛИНЕЙНАЯ МОДЕЛЬ ПЛАНИРОВАНИЯ

# С ГАУССОВСКИМ КРИТЕРИЕМ ОПТИМАЛЬНОСТИ

**(Модель Гауссовского программирования)**

Михаил Любощинский

*Мы предлогаем Экономико – Вероятностную аналогию: категория Стоимости подобна категории Вероятности. Предложенная аналогия позволяет построить содержательную нелинейную невыпуклую Гауссовскую Теорию Полезности и Стоимости, которая более адекватно описывает реальные экономические процессы, чем теории, основанные на линейных и выпуклых моделях.*

*На основе предложенной аналогии, мы конструируем нелинейную невыпуклую модель планирования с Гауссовским критерием оптимальности – модель Гауссовского Программирования.*

*Мы также описываем соответствующую модель Обобщённого Кусочно-Линейного Программирования (ОКЛП) которая может быть использована для аппроксимации модели Гауссовского Программирования, и наоборот.*

*Предлагаемые конструкции продемонстрированы на численном примере.*

## I. О ЯЗЫКЕ ЭКОНОМИЧЕСКОГО ВЗАИМОДЕЙСТВИЯ, УЧИТЫВАЮЩЕГО ОСОБЕННОСТИ РЕАЛЬНЫХ ЭКОНОМИЧЕСКИХ СИСТЕМ.

Процессы планирования, в реальных системах, часто отличаются такими специфическими чертами, которые трудно одновременно и просто, без нагромождения, учесть в рамках существующих экономико-математических моделей. Например, машиностроительные, химические, и многие другие предприятия находятся в условиях комплектного планирования, когда каждый из заказов представляет собой комплект связанных продуктов. При этом, наиболее выгодным является точное выполнение принятых заказов, так как при невыполнении плана, хотя бы по одному продукту, входящему в комплект, невыполненным считается весь заказ, и это не компенсируется перевыполнением объемов по другим продуктам.

Кроме того, на внутризаводском уровне, помимо комплектности, специфика производства, технологии, часто характеризуется «дискретностью» требований по ресурсам (либо столько, сколько нужно, либо ничего): для выпуска единицы продукции по рецептурному составу требуется вполне определенный набор сырья в определенных пропорциях и количествах. При отсутствии одной из компонент или недостатке, в количестве, превышающем заданный допуск на отклонение, не нужны и прочие компоненты.

Вообще говоря, взаимозависимость и существование допуска на отдельные компоненты плана - явление, общее для экономических систем любого уровня: практически любая плановая цифра допускает некоторые отклонения в ту или иную сторону. Однако, как правило, величина допуска не формализована, а подразумевается.

Ситуации, аналогичные рассмотренным, можно найти в любой отрасли экономики и на любом уровне планирования. Очевидно, что, как недовыполнение, так и перевыполнение плановых требований, ведут к нарушению баланса в системе. Однако, принятые линейные критерии оценки деятельности, подсчитанные в линейных ценах, не зависящих от количества или объема, непосредственно этого не учитывают, не отражают и не стимулируют точного выполнения плановых требований, и, как в экономической практике, так и в экономико-математических моделях, часто дополняются доплатами, штрафами, премиями, страховками, приплатами и скидками, поправками к ценам и прочими оценочными механизмами. Одновременное использование линейных цен и поправок к ним, приводит к тому, что реальные плановые решения принимаются в условиях фактически нелинейных цен и критериев оценки деятельности.

Практически, с точки зрения экономического субъекта, субъективная полезность $P(x)$, (или стоимость, как численная мера полезности), количества $x$ некоторого ресурса, необходимого для производства, изменяется следующим нелинейным образом, и представляет собой не выпуклую функцию (Рис. 1):

- В диапазоне от нуля до некоторого объема А1 субъективная полезность количества $x$ некоторого ресурса равна 0, т.к. этого объема совершенно не достаточно для удовлетворения технологической потребности в данном ресурсе.
- От объема А1, полезность начинает возрастать до некоторого объема А2.
- В диапазоне от объема А2 и далее, субъективная полезность более не возрастает, т.к. объем А2 уже полностью достаточен для удовлетворения рассматриваемой технологической потребности.

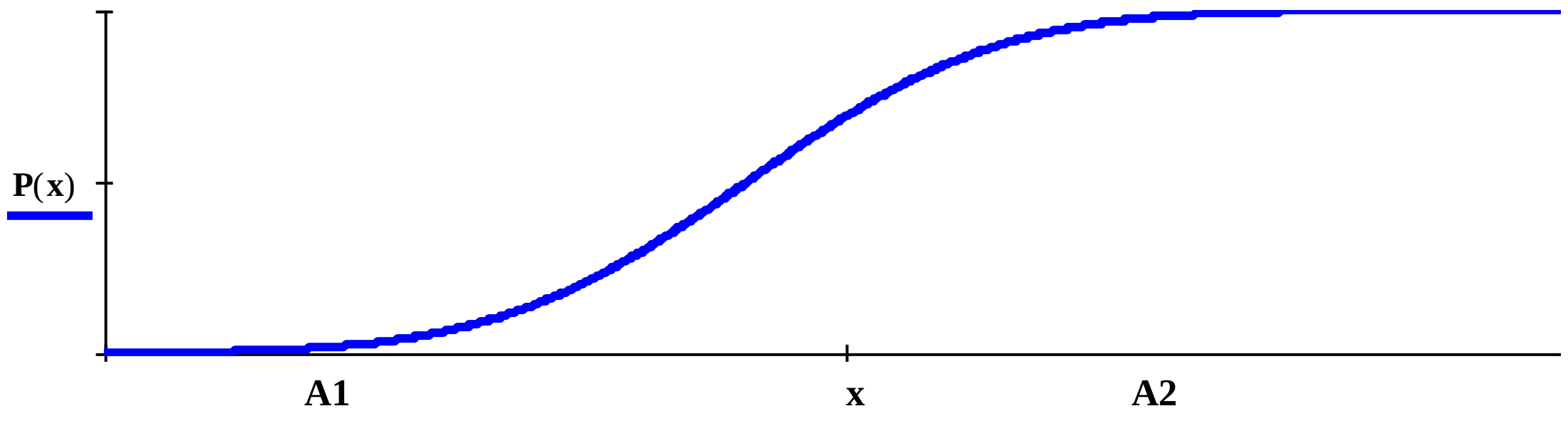

**Рис. 1. Зависимость субъективной полезности P(x) от объёма x продукта.**

Таким образом, требование выпуклости оценочных критериев в экономико-математических моделях, понятное математически, как достаточное для существования и единственности решения, не представляется достаточно общим и обоснованным содержательно-экономически.

**Замечание 1.** Приведённые выше рассуждения относительно неадекватности использования линейных и выпуклых функций для описания стоимости или полезности, полностью применимы и относительно производственной функции, выражающей количественную зависимость между величинами выпуска (количеством продукции) и факторами производства, (затраты ресурсов). Действительно, на микроэкономическом уровне, для очень многих технологий, производство возможно только в некотором диапазоне [A1,A2] наличия ресурса (ресурсов) (Рис.1). В диапазоне [0,A2] – производство невозможно, а в диапазоне [A2,+∞] увеличение количества ресурса не приводит к дальнейшему увеличению объёма выпуска продукции, из-за ограниченности производственной мощности. Так что, производственная функция, так же, принципиально невыпукла!

**<u>Замечание 2.</u>** Вообще говоря, функция полезности описывает спрос, а производственная функция описывает предложение. Так что, на микроэкономическом уровне кривые спроса и предложения, так же, принципиально невыпуклы. На макроэкономическом уровне, этот эффект сглаживается, и создаётся впечатление достаточной адекватности использования линейных или выпуклых функций.

Использование линейных критериев в планировании, кроме того, не обеспечивает достаточной экономической управляемости систем нижнего уровня с точки зрения верхнего уровня: при использовании в процессе согласования локальных моделей, типа моделей линейного программирования, очень часто, малые изменения исходных данных вызывают существенные изменения решения (в случае ортогональности вектора цен одной из граней множества допустимых планов, при незначительных колебаниях вектора цен, решение будет резко перебрасываться из одной вершины симплекса в другую).

Таким образом, используемый «язык» экономического взаимодействия и экономико-математического моделирования, то есть формализация таких понятий, как стоимость, цена, план, спрос, предложение, комплектность, дискретность, точность, штраф, премия, критерий, ограничение, и т.д., и требование выпуклости целевых функций, не соответствуют сути экономических процессов и не позволяет целостно, связанно, одновременно и достаточно просто отразить и учесть перечисленные особенности реальных экономических систем.

Трудности в совершенствовании экономико-математических методов планирования, в разработке адекватных моделей и их практического применения, на наш взгляд, во многом, обусловлены именно неадекватностью языка, в котором они формулируются.

Приведенное на Рис. 1 качественное описание субъективной функции полезности, (или стоимости) $P(x)$, чрезвычайно напоминает функцию нормального распределения вероятности Гаусса. Попытаемся расширить информативность языка экономического взаимодействия, используя, для формализации одного из основных экономических понятий – «план», количественные зависимости, аналогичные зависимостям теории вероятностей.

**<u>Определение 1.</u> В данной работе, План, как совокупность требований к продукции, предлагается формализовать как оценочную структуру, с помощью многомерной ненормированной совместной Гауссовской нормальной плотности распределения стоимости на пространстве продуктов и ресурсов. При таком подходе, Стоимость аналогична Вероятности, а Цена – Плотности Вероятности. При этом, математическое ожидание представляет собой плановые требования по объёмам отдельных компонет плана, а матрица дисперсий – описывает требуемые точности выполнения отдельных компонент плана, и их взаимозависимость.**

**<u>Замечание 3.</u>** Использование Гауссовского распределения для формализации производственной функции и вообще для формализации спроса и предложения на микро и макро экономических уровнях, требует специального рассмотрения.

**<u>Замечание 4.</u>** Можно использовать и Логистическую функцию, но этот вопрос, так же, требует специального рассмотрения.

## <u>Рассмотрим однопродуктовый случай.</u>

**<u>Определение 2.</u>** Определим стоимость объёма поставки $x$ некоторого продукта через функцию $F(x,m,\sigma,\lambda)$, (Гауссовский критерий оптимальности):

**(1)** $$F(x,m,\sigma,\lambda)=\int_0^x f(x,m,\sigma,\lambda)dx$$ , где

**(2)** $$f(x,m,\sigma,\lambda)=\frac{2\lambda}{\sigma\sqrt{2\pi}}\exp\left[\frac{-1}{2}\left(\frac{x-m}{\sigma}\right)^2\right]$$ - плотность распределения стоимости, или нелинейная зависимость цены продукта как функции от объёма $x$ поставки продукта.

В отличие от теоретико-вероятностной функции нормального распределения, интегрирование в (1) производится только в неотрицательной области, имеющей экономический смысл.

Здесь:

$m$ - желаемый для верхнего уровня объем производства продукта,

x – предлагаемый подсистемой объём производства (поставки) данного продукта,

σ - дисперсия, отражающая точность плановых требований по продукту,

λ - предлагаемая стоимость плана, при его точном выполнении (x = m).

**Замечание 5.** Максимальная стоимость плана, при его переыполнении более чем на $m+3\sigma$, не превосходит $2\lambda$, т.е $F(x) < 2\lambda$. Таким образом, критерий оптимальности $F(x)$ содержит в себе и штраф за «недовыполнение» плана поставки $m$, и премию за «перевыполнение».

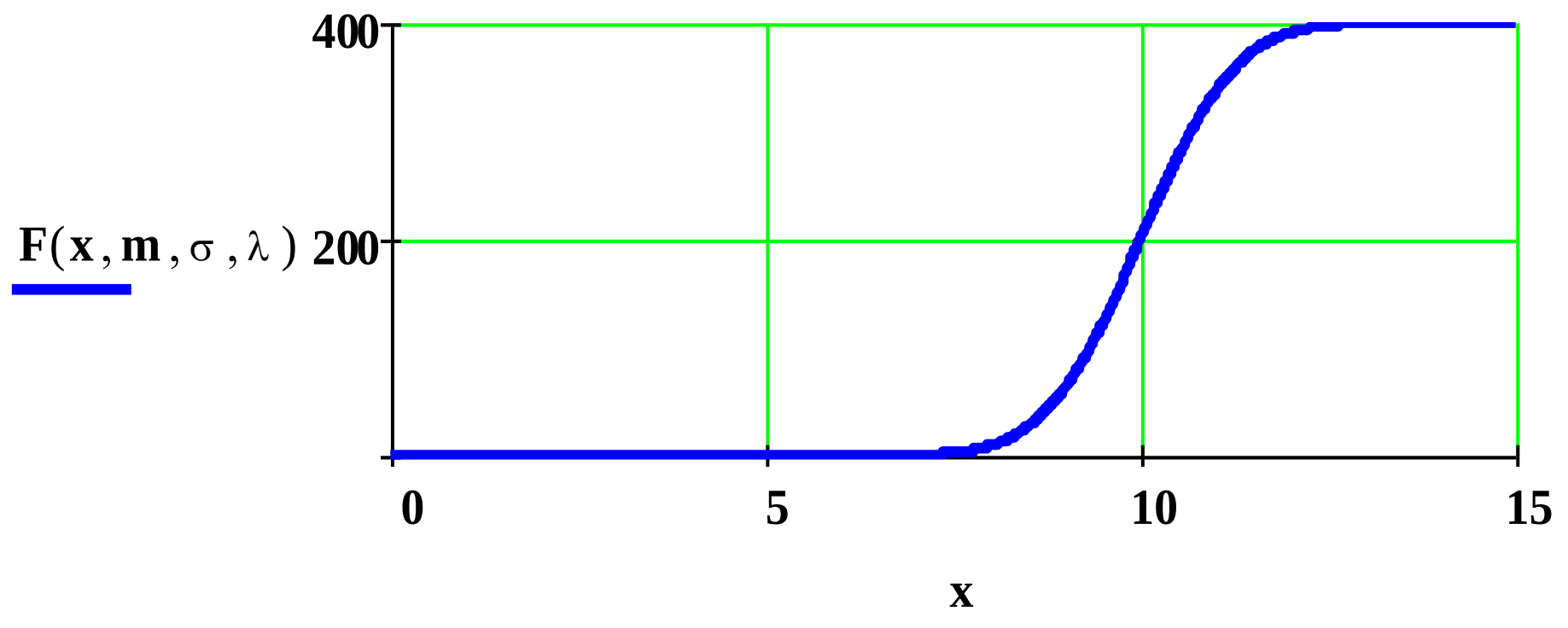

(Функция распределения стоимости)

**Рис. 2. Зависимость стоимости продукта от количества *x* поставки продукта при плане m=10, допустимой точности поставки σ=1, и «точной» стоимости λ=200.**

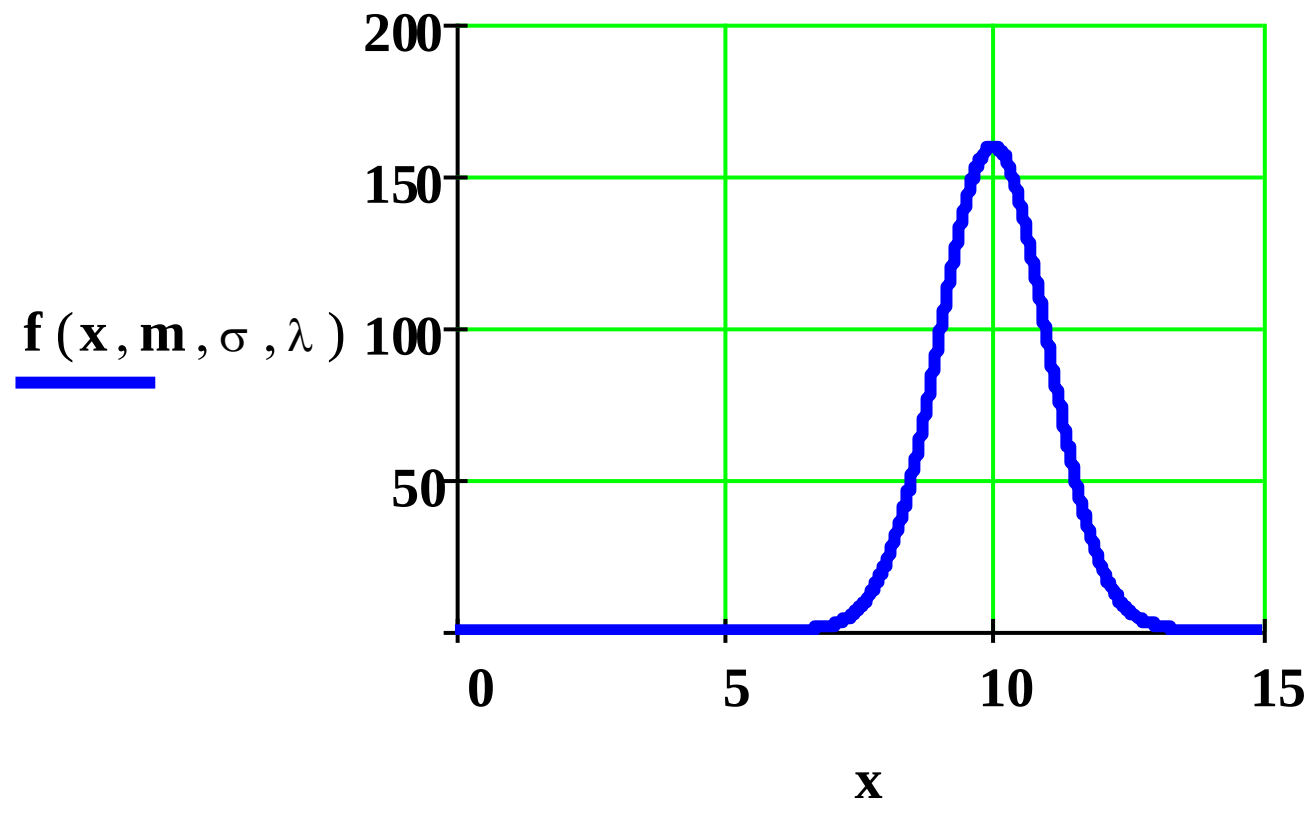

(Цена, как плотность распределения стоимости)

**Рис. 3. Зависимость цены единицы продукта от количесыва продукта при плане m=10, допустимой точности поставки σ=1, и «точной» стоимости λ=200.**

## Перейдём к многопродуктовому случаю.

**а. В случае некомплектных требований по продуктам,** определим стоимость многопродуктового вектора поставки $\bar{x}$ взаимонесвязанных продуктов $x_i$ следующим образом:

**(3)** $$F_н(\bar{x},\bar{m},\bar{\sigma},\lambda)=\sum_i F_i(x_i,m_i,\sigma_i,\lambda_i)=\sum_i \int_0^{x_i} f_i(x_i,m_i,\sigma_i,\lambda_i)dx_i$$

где $\lambda=\sum_i \lambda_i$ - «точная» стоимость плана .

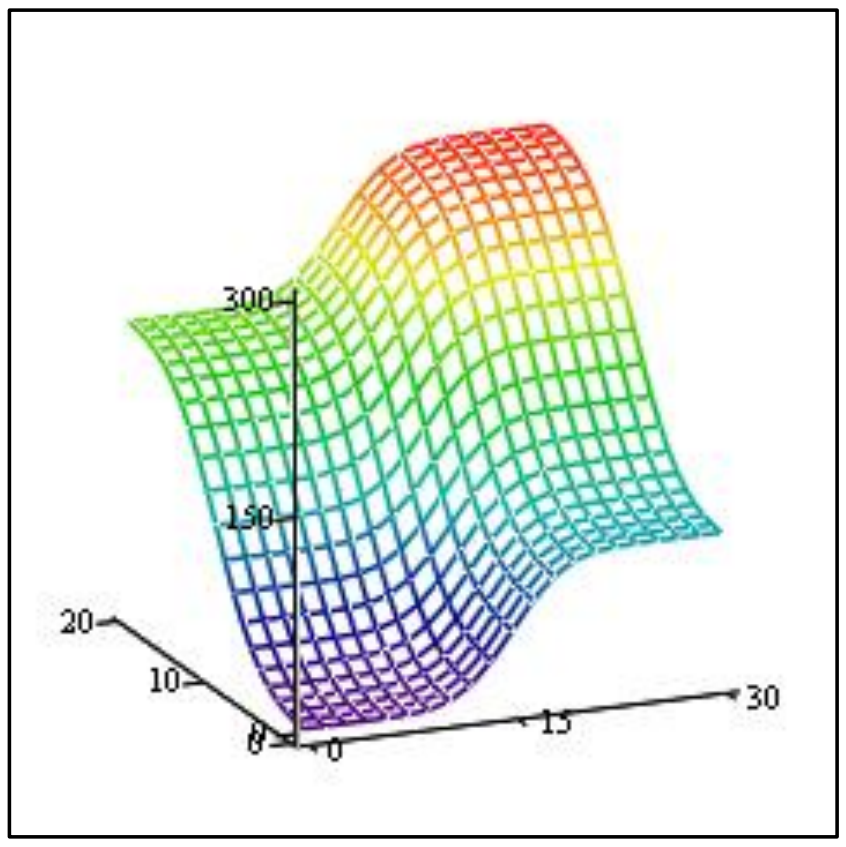

F
График

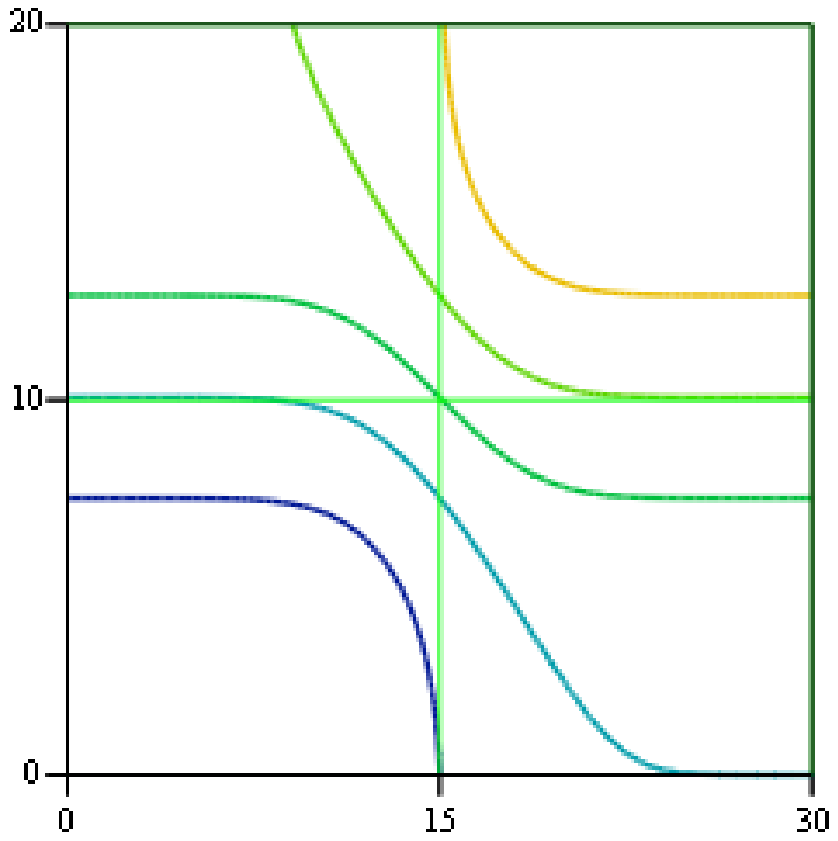

F
Кривые безразличия

Функции распределения стоимости поставки двух комплектно независимых продуктов.

**Рис. 4. Зависимость стоимости от количества (*x,y*) поставки двух некомплектных продуктов, при плане** $\bar{m}=(15{,}10)$**, допустимой точности поставки** $\bar{\sigma}=(3{,}4)$**, «точной» стоимости компонент** $\bar{\lambda}=(50{,}100)$**, и «точной» стоимости всего плана** $\lambda=150$ **.**

**б. Для взаимосвязанных, комплектных, продуктов,** определим стоимость многопродуктового вектора-комплекта поставки $\bar{x}$ как многомерную функцию нормального распределения Гаусса:

**(4)** $$F_з(\bar{x},\bar{m},\Sigma,\lambda)=\iiint_0 f(\bar{x},\bar{m},\Sigma,\lambda)d\bar{x}$$ , где

**(5)** $$f(\bar{x},\bar{m},\Sigma,\lambda)=\frac{2\lambda}{(2\pi)^{\frac{n}{2}}|\Sigma|^{\frac{1}{2}}}\exp[-\frac{1}{2}(\bar{x}-\bar{m})\Sigma^{-1}(\bar{x}-\bar{m})^T]$$

где $\lambda$ - «точная» стоимость комплектного плана при точном выполнении $\bar{x} = \bar{m}$ плана, $\Sigma$ – ковариационная матрица, характеризующая точности и взаимозависимость отдельных компонент плана, n – число продуктов в комплекте. Таким образом, помимо комплектности продуктов, можно учесть и их корреляцию.

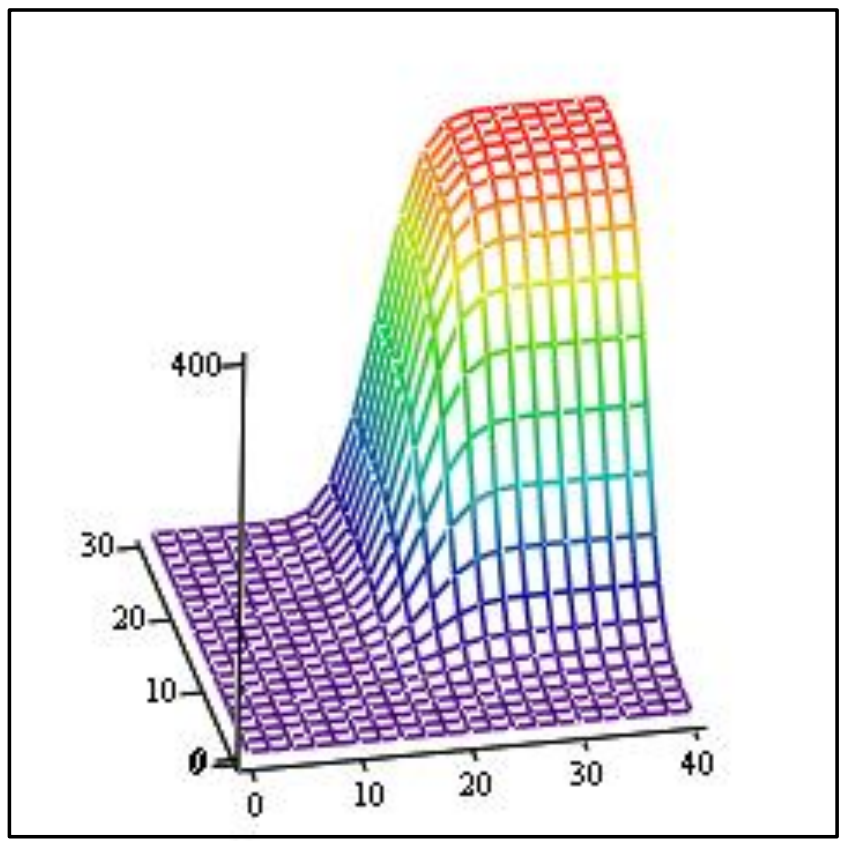

F

График функции

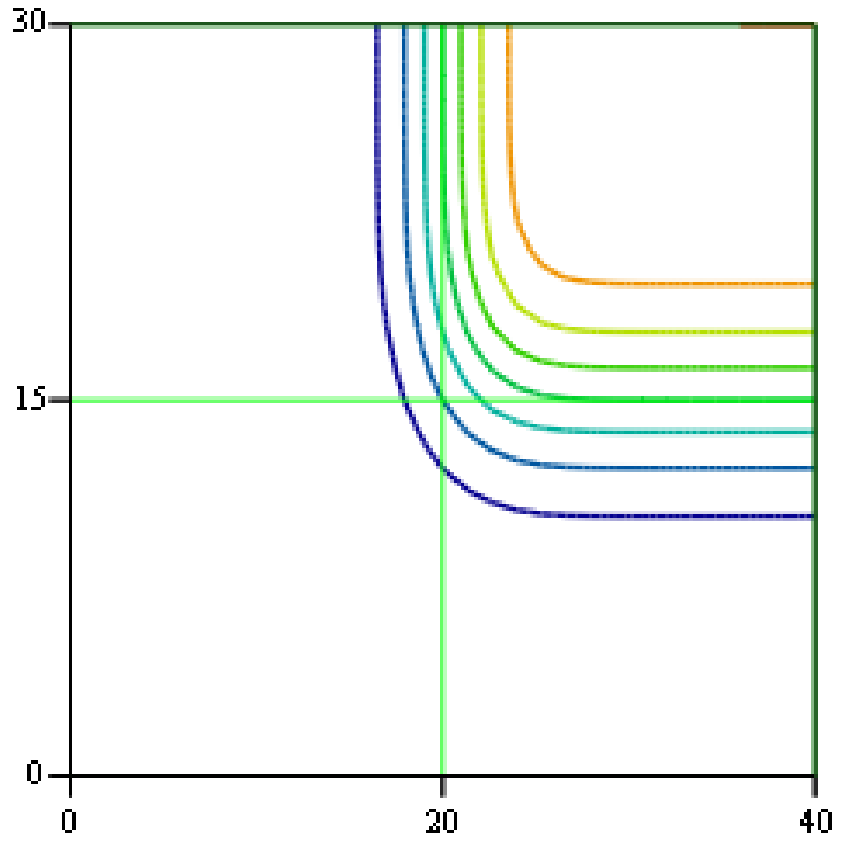

F

Кривые безразличия

<u>Функции распределения стоимости *F(x,y)* поставки двух комплектных продуктов</u>.

**<u>Рис. 5. Зависимость стоимости *F* от количества (*x,y*) поставки двух комплектных продуктов, при плане</u> $\bar{m} = (20{,}15)$, <u>допустимой точности поставки</u> $\bar{\sigma} = (3{,}4)$, <u>и «точной» стоимости комплекта λ=200</u>.**

**<u>в. В смешанном случае, при наличии комплектных и некомплектных продуктов</u>**, определим суммарную «совместную» стоимость многопродуктового вектора поставки, как сумму стоимостей всех зависимых и независимых компонент плана:

**(6)** $$F(\bar{x}) = \sum F_з(\bar{x}) + \sum F_н(\bar{x})$$

Предложенный критерий оптимальности $F(\bar{x})$, построенный таким образом, является строго монотонно-возрастающей по всем переменным функцией, что, вообще говоря, не достаточно для обеспечения единственности решения оптимизационной задачи с таким критерием.

Вопросы существования и единственности решения, а так же разработки методов и алгоритмов решения задач Гауссовского планирования, требуют специального рассмотрения, и вынесены за рамки данной работы.

## II. МОДЕЛЬ НЕЛИНЕЙНОГО ПЛАНИРОВАНИЯ

## С ГАУССОВСКИМ КРИТЕРИЕМ ОПТИМАЛЬНОСТИ

### (МОДЕЛЬ ГАУССОВСКОГО ПРОГРАММИРОВАНИЯ)

Рассмотрим следующую задачу нелинейного программирования с Гауссовской функцией распределения стоимости $F(\bar{x}, \bar{m}, \bar{\sigma}, \lambda)$ и с линейным оператором затрат $A(\overline{\mathrm{x}})$ – модель Гауссовского программирования:

**(7)**
$$\begin{aligned} &F(\bar{x}, \bar{m}, \Sigma, \lambda) \Rightarrow Max_{\bar{\mathrm{x}}} \\ &A \cdot \bar{x} \le \bar{r} \\ &\bar{x} \ge 0, \end{aligned}$$

Где:

$F(\bar{x}, \bar{m}, \Sigma, \lambda)$ - Гауссовская функция распределения стоимости на пространстве продуктов,

$\bar{x}$ – вектор производства продуктов,

А – матрица затрат ресурсов,

$\bar{r}$ – вектор наличия ресурсов.

Рассмотрим соответствующую задаче (7) функцию Лагранжа:

**(8)**
$$L(\bar{x}, \bar{y}) = F(\bar{x}, \bar{m}, \Sigma, \lambda) - \bar{y}A\bar{x} + \bar{y}\bar{r}$$

Выпишем условия Куна-Таккера [1] для функции Лагранжа (8):

**(9)**
$$\begin{aligned} &\overline{\nabla}_{\bar{x}} L(\bar{x}, \bar{y}) = \bar{g}(\bar{x}) - \bar{y}A \le 0, \qquad \text{где} \quad \bar{g}(\bar{x}) = \overline{\nabla}_{\bar{x}} F(\bar{x}, \bar{m}, \Sigma, \lambda) \\ &\overline{\nabla}_{\bar{y}} L(\bar{x}, \bar{y}) = \bar{r} - A\bar{x} \ge 0 \\ &(\bar{g}(x) - \bar{y}A)\bar{x} = \bar{0} \qquad | \quad \text{Условия дополняющей нежёсткости.} \\ &\bar{y}(\bar{r} - A\bar{x}) = \bar{0} \qquad | \\ &\bar{x} \ge 0 \qquad \bar{y} \ge 0 \end{aligned}$$

На основе (8) и (9) составим двойственную к (7) задачу:

$$\bar{y}\bar{r} \Rightarrow Min_{\bar{y}}$$

**(10)** $$\bar{y}A \ge \bar{g}(\bar{x})$$

$$\bar{y} \ge 0$$

**<u>Замечание 6.</u>** Необходимо чётко различать два вида стоимостных категорий в некоторой точке $\bar{x}'$ - градиентные (локальные) и интегральную.

**<u>Определение 3.</u>** « Градиентная Цена плана» в точке $\bar{x}'$:

**(11)** $$\bar{g}(\bar{x}') = \nabla_{\bar{x}} F(\bar{x}, \bar{m}, \Sigma, \lambda)\big|_{x'}$$

**<u>Определение 4.</u>** «Градиентная Стоимость плана» в точке $\bar{x}'$:

**(12)** $G(\bar{x}') = \bar{g}(\bar{x}') \cdot \bar{x}'$ - стоимость плана $\bar{x}'$, посчитанная на основе линейной Градиентной Цены (локальная стоимостная оценка плана $\bar{x}'$).

**<u>Определение 5.</u>** «Интегральная Стоимость плана», или «Полная Стоимость плана» $\bar{x}'$ - значение нелинейного критерия $F(\bar{x}, \bar{m}, \Sigma, \lambda)\big|_{x'}$ в точке $\bar{x}'$ (глобальная характеристика).

**<u>Замечание 7.</u>** Градиентная Цена $\bar{g}(\bar{x}')$ и Градиентная Стоимость $G(\bar{x}')$, по сути, являются маргинальными оценками плана $\bar{x}'$, характеризующими изменение критерия оптимальности $F(\bar{x})$ (Полной Стоимости) при незначительных изменениях плана $\bar{x}'$.

Пусть $(\bar{x}^*, \bar{y}^*)$ - точка оптимума задач (7) и (10), где: $\bar{x}^*$ - вектор оптимального плана производства, а $\bar{y}^*$ - вектор оптимальных внутренних цен (двойственных оценок) на ресурсы.

Тогда из условий дополняющей нежёсткости в (9) получаем следующие балансовые соотношения (13) – (15), имеющих прозрачный экономический смысл:

**(13)** $\bar{g}(\bar{x}^*)\cdot\bar{x}^* = \bar{y}^*\cdot(A\cdot\bar{x}^*)$

- Полная градиентная (маргинальная) стоимость произведённой продукции во внешних градиентных ценах в точке $\bar{x}^*$, равна полной внутренней стоимости израсходованных ресурсов $A\cdot\bar{x}^*$ во внутренних ценах $\bar{y}^*$.

**Таким образом, внутренняя стоимость ресурсов определяется внешней градиентной, или маргинальной, стоимостью продукции в точке $\bar{x}^*$. (Связь внутренних и внешних стоимостных категорий).**

**(14)** $\bar{y}^*\cdot(A\cdot\bar{x}^*) = \bar{y}^*\cdot\bar{r}$

- Полная внутренняя стоимость израсходованных ресурсов, равна полной внутренней стоимости, начального запаса ресурсов $\bar{r}$. (Это соотношение выполняется и покомпонентно, для каждого ресурса, и агрегированно).

Из соотношений (12), (13) следует:

**(15)** $\bar{g}(\bar{x}^*)\bar{x}^* = \bar{y}^*\bar{r} = \bar{y}^*(A\bar{x}^*)$

- Полная внешняя градиентная стоимость произведённой продукции $\bar{g}(\bar{x}^*)\cdot\bar{x}^*$ равна полной внутренней стоимости $\bar{y}^*\cdot\bar{r}$ (в двойственных линейных ценах $\bar{y}^*$) начального запаса ресурсов $\bar{r}$, или израсходованных ресурсов $A\cdot\bar{x}^*$. ( Расширение соотношений (13) и (14)).

**Замечание 8.** Как известно [1], в модели линейного программирования, в точке оптимума, совпадают значения критериев прямой и двойственной задач, т.е. полная внешняя стоимость произведённой продукции равна полной внутренней стоимости израсходованных (или начальных) ресурсов.

**В модели Гауссовского планирования, в точке оптимума выполняется баланс между полной внешней градиентной (маргинальной) стоимостью произведённой продукции и полной внутренней стоимостью израсходованных (или начальных) ресурсов. Внешняя полная стоимость $F(\bar{x}^*,\bar{m},\bar{\sigma},\lambda)$ плана $\bar{x}^*$, вообще говоря, не равна полной внутренней**

**стоимости израсходованных (или начальных) ресурсов $\bar{y}^*\cdot\bar{r}$: при росте $\bar{x}^*$, $F(\bar{x}^*,\bar{m},\bar{\sigma},\lambda)$ будет приближаться к максимальной стоимости $2\lambda$ плана , а маргинальная (градиентная) стоимость $\bar{g}(\bar{x}^*)\cdot\bar{x}^*$ и, равная ей, внутренняя стоимость ресурсов, будут, вначале возрастать ( до т. $\bar{x}^*=\bar{m}$), а затем уменьшаться, приближаясь к 0.**

**Замечание 9.** Модель (6) очень интересна с практической т.зрения, т.к. имеет прямую связь с теорией экспертных оценок: Гауссовский критерий $F(x,m,\sigma,\lambda)$ интуитивно значительно понятнее, чем линейный критерий $\bar{c}\cdot\bar{x}$, где $\bar{c}$ - вектор цен на продукцию.

Дело в том, что эксперту легче оценить и представить информацию о необходимом объёме поставки $m$ , требуемой точности поставки $\sigma$ , взаимосвязанности отдельных ресурсов и стоимости точной поставки $\lambda_i$ отдельных несвязанных ресурсов (или комплектов в целом, без оценки стоимости отдельных компонент), чем только информацию о цене (как в модели ЛП), т.к. три параметра - объём, точность и стоимость экономически и интуитивно очень сильно связаны.

В модели линейного программирования этой связи, очевидно, нет: в критерии $\bar{c}\cdot\bar{x}$ прямой задачи фигурирует только цена (а не стоимость), а объём $m$, точность $\sigma$ и комплектность - вообще "за кадром"! Это, отчасти, объясняет трудности применения модели ЛП в экономике.

## III. Связь моделей Гауссовского и Линейного планирования.

Пусть $f(x,m,\sigma)$ - Гауссовская функция плотности распределения вероятности случайной величины $x$, с математическим ожиданием $m$ и дисперсией $\sigma^2$.

Рассмотрим $u(x,a,b)$ – функцию плотности распределения вероятности случайной величины $x$, равномерно распределённой на интервале [a,b]. Как известно [2], её плотность $u$, математическое ожидание $m$ и дисперсия $\sigma^2$, определяются следующими соотношениями:

**(16)**
$$u=\frac{1}{a+b}$$
$$m=\frac{a+b}{2}$$
$$\sigma=\frac{b-a}{2\sqrt{3}}$$

Из (16) выразим $a$ и $b$ через $\sigma$ и $m$:

$$a = m - \sigma\sqrt{3}$$

**(17)**

$$b = m + \sigma\sqrt{3}$$

Сопоставим графики функций распределения вероятностей $F(x,m,\sigma) = \int_{-\infty}^{x} f(x,m,\sigma)dx$ и $U(x,a,b) = \int_{-\infty}^{x} u(x,a,b)dx$ , где $f(x,m,\sigma)$ и $u(x,a,b)$ - функции распределения плотности вероятности Гауссовского и равномерного распределений, соответственно, а интервал [a,b] определяется по формулам (17) .

При $m = 80$ и $\sigma = 20$, имеем: $a = 45.4$, а $b = 114.6$.

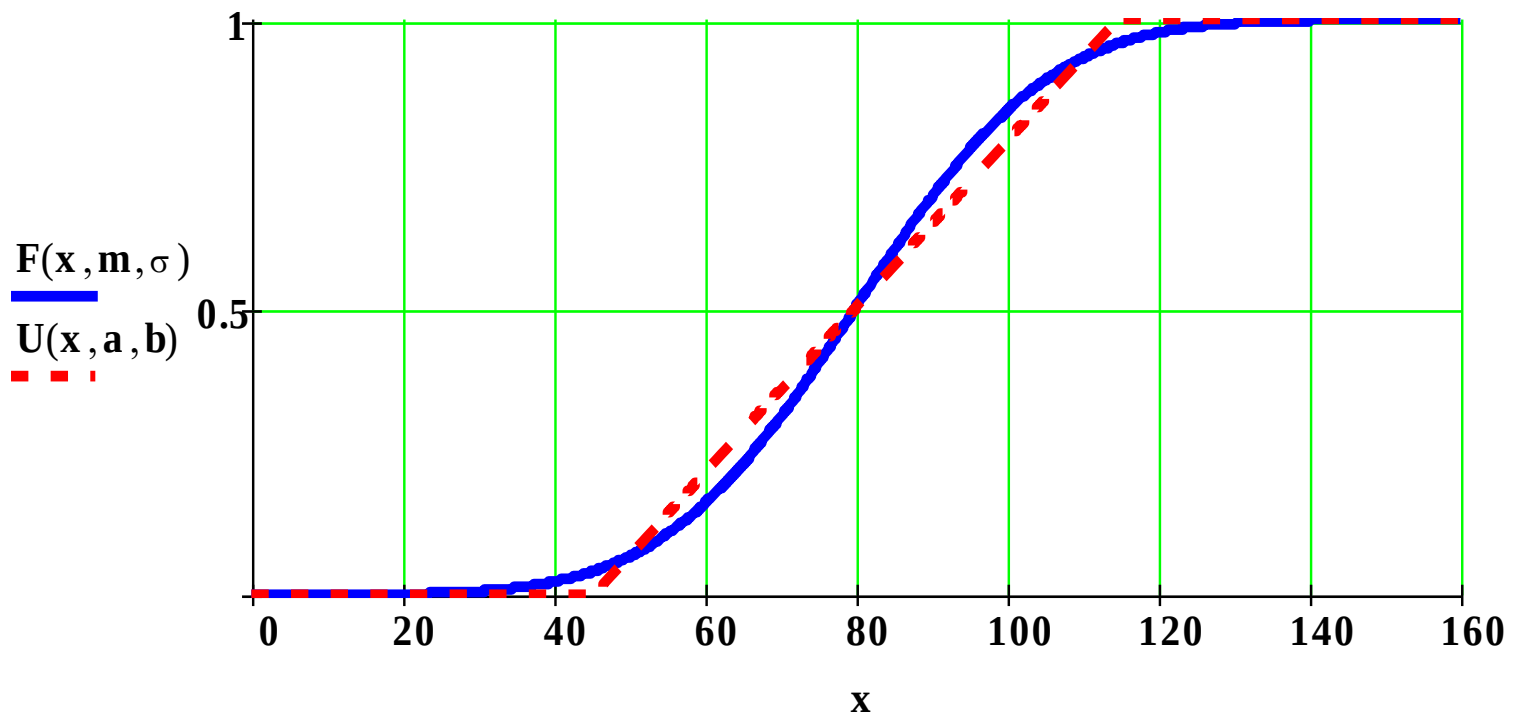

**a m b**

**Рис. 6. Графики функций распределения *F(x,80,20)* и *U(x,45.4,114.6)*.**

Как видно из **Рис. 6**, функции *U(x,45.4,114.6)* и *F(x,80,20)* достаточно хорошо аппроксимирует друг друга.

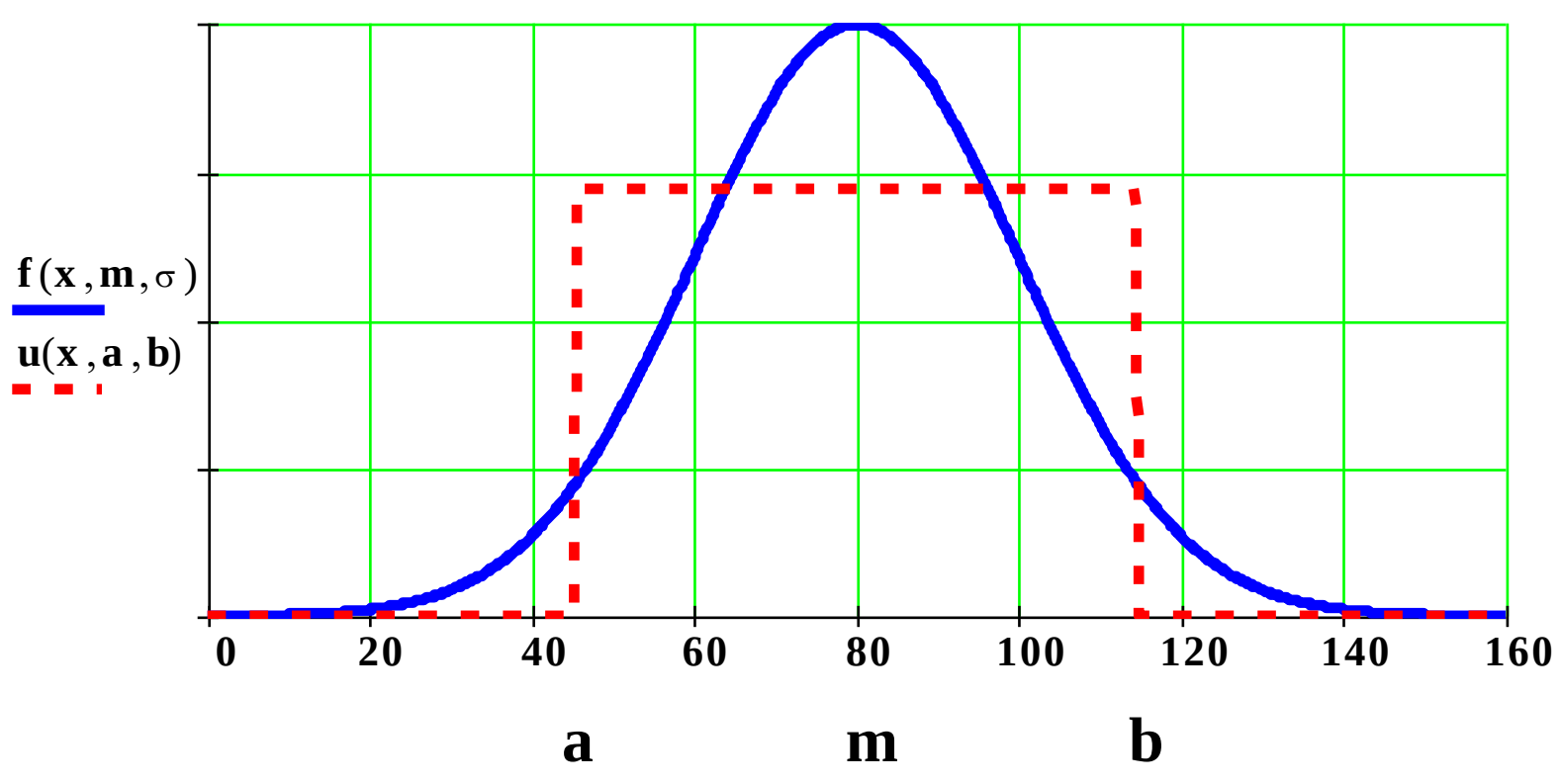

**Рис. 7. Графики функций плотности распределений *f(x,80,20)* и *u(x,45.4,114.6).***

## Гауссовская аппроксимация модели Линейного программирования

Пусть дана модель Линейного программирования:

**(18)**
$$\begin{cases} \overline{cx} \Rightarrow \max_{\overline{x}} \\ A\overline{x} \le \overline{r} \\ \overline{x} \ge 0 \end{cases}$$

Для каждой переменной $x_j$, $j = \overline{1,n}$ , где $n$ - число продуктов, определим интервал допустимых значений $[a_j, b_j]$ таким образом, чтобы область допустимых значений, задаваемая системой ограничений $\{A\overline{x} \le \overline{r}, \overline{x} \ge 0\}$ полностью принадлежала многомерному прямоугольному парралелепипеду $B = \{\overline{x} | x_j \in [a_j, b_j], j = \overline{1,n}\}$:

**(19)**
$$a_j = 0,\ j = \overline{1,n}$$

$$b_j = \max_i \frac{r_j}{a_{ij}},\quad \text{i}= \overline{1,m},$$ где $m$- число ресурсов, $a_{ij} \ne 0$ – элементы матрицы $A$.

Рассмотрим ненормированную функцию плотности равномерного распределения стоимости на интервале $[0, b_j]$:

**(20)**
$$u_j(x_j, 0, b_j, c_j) = \begin{cases} c_j, & 0 \le x_j \le b_j \\ 0, & b_j < x_j \end{cases}$$ , где $c_j$ – цена продукта $x_j$.

Тогда функция равномерного распределения стоимости будет иметь следующий вид:

**(21)** $$U_j(x_j, 0, b_j, \lambda_j) = \begin{cases} c_j x_j, & 0 \le x_j \le b_j \\ c_j b_j, & b_j < x_j \end{cases}$$

В соответствии с (16) определим параметры $u_j, m_j, \sigma_j$:

**(22)** $$\lambda_j = \frac{c_j}{2}$$ $$m_j = \frac{b_j}{2}$$ $$\sigma_j = \frac{b_j}{2\sqrt{3}}$$

Определим Гауссовскую функцию плотности распределения стоимости по каждой переменной $x_j$ следующим образом:

**(23)** $$f_j(x_j, m_j, \sigma_j, \lambda_j) = \frac{2\lambda_j}{\sigma_j\sqrt{2\pi}} \exp\left[\frac{-1}{2}\left(\frac{x_j - m_j}{\sigma_j}\right)^2\right]$$

Тогда функция распределения стоимости по каждой компоненте будет иметь вид:

**(24)** $$F_j(x_j, m_j, \sigma_j, \lambda_j) = \int_0^x f_j(x, m_j, \sigma_j, \lambda_j)dx$$

На **Рис.8** представлены графики функций распределений стоимости, равномерной (линейной) $U_j(x_j, 0, b_j, c_j)$, и Гауссовской $F_j(x_j, m_j, \sigma_j, \lambda_j)$, по каждой компоненте:

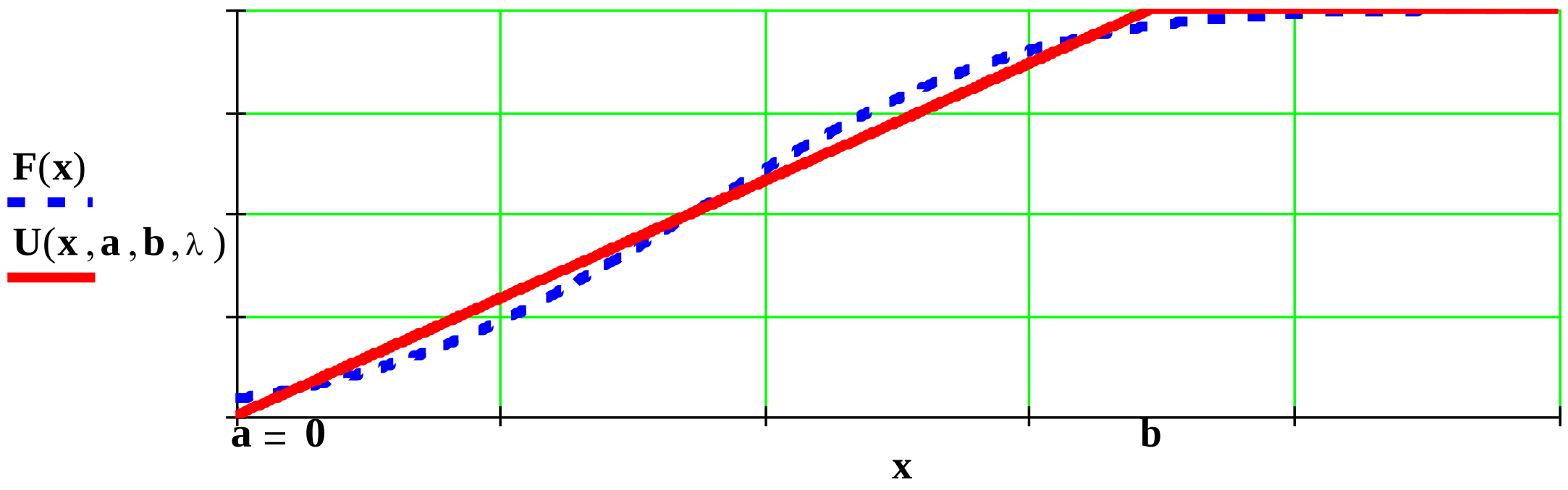

**Рис. 8. Графики функций распределений стоимости $F_j(x_j, m_j, \sigma_j, \lambda_j)$ и $U_j(x_j, 0, b_j, c_j)$.**

Суммарную функцию распределения стоимости определим следующим образом:

**(25)** $$F(\overline{x},\overline{m},\overline{\sigma},\lambda) = \sum_{j=1}^{n} F_j(x_j,m_j,\sigma_j,\lambda_j), \quad \text{где } \lambda = \sum_{j=1}^{n} \lambda_j.$$

Тогда искомая Гауссовская аппроксимация задачи линейного программирования (18) имеет следующий вид:

**(26)** $$\begin{cases} F(\overline{x},\overline{m},\overline{\sigma},\lambda) \Rightarrow \max_{\overline{x}} \\ A\overline{x} \le \overline{r} \\ \overline{x} \ge 0 \end{cases}$$

## Линейная аппроксимация модели Гауссовского планирования

Рассмотрим задачу Гауссовского планирования с некомплектными (независимыми) требованиями по продуктам:

**(27)** $$\begin{cases} F(\overline{x},\overline{m},\overline{\sigma},\lambda) \Rightarrow \max_{\overline{x}} \\ A\overline{x} \le \overline{r} \\ \overline{x} \ge 0 \end{cases},$$

где Гауссовский критерий оптимальности $F(\overline{x},\overline{m},\overline{\sigma},\lambda)$ определяется соотношениями (25), (24) и (23), т.е., определяется набором покомпонентных функций плотности распределения стоимости независимых продуктов $f_j(x_j,m_j,\sigma_j,\lambda_j), j = \overline{1,n}$.

Каждой Гауссовской функции плотности распределения стоимости $f_j(x_j,m_j,\sigma_j,\lambda_j)$, продукта $j$, поставим в соответствие ненормированную функцию плотности равномерного распределения стоимости $u_j(x_j,a_j,b_j,\lambda_j)$, где параметры $a_j, b_j$ определяются по формулам:

**(28)** $$a_j = m_j - \sigma_j\sqrt{3}$$
$$b_j = m_j + \sigma_j\sqrt{3}$$

$$u_j(x_j,a_j,b_j,\lambda_j) = \begin{cases} \frac{2\lambda_j}{b_j-a_j}, & \text{при } x \in [a_j,b_j] \\ 0, & \text{при } x \notin [a_j,b_j] \end{cases}$$

Очевидно, что $U_j(x_j, a_j, b_j, \lambda_j)$ – функция равномерного распределения стоимости продукта $x_j$, на интервале $[a_j, b_j]$ определяется соотношением:

**(29)** $$U_j(x_j, a_j, b_j, \lambda_j) = \begin{cases} \frac{2\lambda_j(x_j - a_j)}{b_j - a_j}, & \text{при } x \in [a_j, b_j] \\ 0, & \text{при } x \notin [a_j, b_j] \end{cases}$$

Определим функцию:

**(30)** $$U(\bar{x}, \bar{a}, \bar{b}, \bar{\lambda}) = \sum_{j=1}^{n} U_j(x_j, a_j, b_j, \lambda_j).$$

Тогда следующая задача с кусочно-линейным критерием оптимальности аппроксимирует исходную задачу (27) Гауссовского программирования:

**(31)** $$\begin{cases} U(\bar{x}, \bar{a}, \bar{b}, \bar{\lambda}) \Rightarrow \max_{\bar{x}} \\ A\bar{x} \leq \bar{r} \\ \bar{x} \geq 0 \end{cases}$$

Модель (29)-(31) представляет собой модель Кусочно-Линейного Программирования и является обобщением классической модели линейного программирования. Её можно назвать моделью Обобщённого Кусочно-Линейного Программирования – ОКЛП.

В классической модели ЛП, цена постоянна на всей области значений каждой переменной $x_j$, в модели (31), цена постоянна на интервале $[a_j, b_j]$, а вне интервала равна 0.

Классическая модель линейного программирования – это модель с равномерным распределением стоимости на пространстве продуктов.

Модель ОКЛП представляет собой модель с равномерным распределением стоимости на многомерном прямоугольном парралелепипеде $[\bar{a}, \bar{b}]$.

**Замечание 10.** Задачу Гауссовского планирования с коррелированными компонентами плана можно аппроксимировать ОКЛП задачей типа (31) при помощи поворота и переноса осей координат, в которых сформулирована исходная задача Гауссовского программирования.

Таким образом, любую задачу Линейного программирования, включая ОКЛП, можно аппроксимировать моделью Гауссовского планирования, и любую задачу Гауссовского планирования можно аппроксимировать моделью обобщённого кусочно-линейного программирования.

Степень адекватности таких аппроксимаций, безусловно, требует специального рассмотрения.

**Замечание 11.** Модель обобщённого кусочно-линейного программирования значительно информативнее классической модели ЛП: она позволяет учесть принципиальную нелинейность и невыпуклость реальных функций полезности, описанных при рассмотрении Рис. 1 в начале данной работы.

**Замечание 12.** При решении практических экономических задач, возможны два подхода:

1) Параметры $\overline{a}, \overline{b}$ и $\overline{\lambda}$ модели (31) определяются на основе экспертных оценок, а модель Гауссовского планирования строится как аппроксимация полученной, на основе экспертных оценок, модели ОКЛП.
2) Параметры $\overline{m}, \Sigma, \lambda$ получаются непосредственно, используя специальную методику экспертных оценок, и формируется модель Гауссовского планирования.

## IV. Числовой пример задачи с 4-мя продуктами: двух независимых и двух "комплектных".

Исходная информация:

$\overline{m} = (30 \quad 40 \quad (900 \quad 100))$ кг. - вектор требуемых объёмов поставки продуктов.

$\overline{\sigma} = (10 \quad 13 \quad (300 \quad 30))$ кг. - вектор требуемых точностей поставки продуктов.

$\overline{\lambda} = (5{,}000.00 \quad 10{,}000.00 \quad 200{,}000.00)$ руб. - вектор стоимости компонентов «точного» плана.

**(32)** $$A = \begin{bmatrix} 0 & 40 & 50 & 30 \\ 30 & 0 & 10 & 0 \\ 70 & 40 & 0 & 20 \end{bmatrix}$$ - матрица затрат ресурсов на производство продуктов.

$\overline{b} = (49{,}500 \quad 9{,}900 \quad 5{,}700)$ кг. - вектор начального запаса ресурсов.

Задача Планирования с нелинейным Гауссовским критерием оптимальности:

$$F(\bar{x}) = 2\lambda_1 \int_0^{x_1} f(x_1, m_1, \sigma_1) dx_1 + 2\lambda_2 \int_0^{x_2} f(x_2, m_2, \sigma_2) dx_2 +$$

$$+ 2\lambda_3 \int_0^{x_3}\int_0^{x_4} f(x_3, m_3, \sigma_3) f(x_4, m_4, \sigma_4) dx_4 dx_3$$

**(33)** $F(\bar{x}) \Rightarrow \max_{\bar{x}}$

$A\bar{x} \le \bar{b}$

$\bar{x} \ge 0$,

где $f(x, m, \sigma)$ - обычная Гауссовская плотность нормального распределения.

(Для простоты рассматривается случай комплектных, но некоррелированных продуктов $x_3$ и $x_4$, с общей стоимостью $\lambda_3$ плана при точном выполнении: $x_3 = m_3$, и $x_4 = m_4$).

Используя существующие пакеты программ для решения задач нелинейного программирования, найдём решение задачи (33) с исходными данными (32).

(В данной работе использовалась система Mathcad):

$$\bar{x}^* = \begin{pmatrix} 33.104 \\ 0 \\ \begin{pmatrix} 890.688 \\ 165.52 \end{pmatrix} \end{pmatrix}$$ - решение нелинейной оптимизационной задачи (33) (или (7))

**(34)**

$F(\bar{x}^*) = 197{,}812.64$ руб. - Полная Стоимость Плана $\bar{x}^*$ (оптимальное значение критерия)

Для решения двойственной задачи к (33), определим, в найденной точке оптимума $\bar{x}^*$, значение вектора «градиентных» цен $\bar{g}(\bar{x}) = \nabla_{\bar{x}} F(\bar{x})$:

**(35)** $$\bar{g}(\bar{x}^*) = \begin{pmatrix} 380.18 \\ 5.4 \\ 523.74 \\ 238.21 \end{pmatrix}$$ руб/кг

Используя (35) и (10), запишем двойственную к (33) задачу:

$$\bar{y}\bar{b} \Rightarrow Min_{\bar{y}}$$

**(36)** $$\bar{y}A \geq \bar{g}(\bar{x}^*) = \begin{pmatrix} 380.18 \\ 5.4 \\ 523.74 \\ 238.21 \end{pmatrix}$$

$$\bar{y} \geq 0$$

Найдём решение этой задачи линейного программирования:

$$\bar{y}^* = \begin{pmatrix} 7.94 \\ 12.67 \\ 0 \end{pmatrix}$$ руб/кг - оптимальное решение двойственной к (33) задачи (36).

**(37)**

$$\bar{y}^* \cdot \bar{b} = 518{,}502.36$$ руб. - Полная стоимость ресурсов во внутренних ценах

(оптимальное значение критерия двойственной задачи.)

Рассмотрим, как при найденном оптимальном плане $\bar{x}^*$, выполняются плановые требования $\bar{m}$ и ограниченя по ресурсам.

$$\bar{m} = \begin{pmatrix} 30 \\ 40 \\ \begin{pmatrix} 900 \\ 100 \end{pmatrix} \end{pmatrix}; \quad \bar{x}^* = \begin{pmatrix} 33.104 \\ 0 \\ \begin{pmatrix} 890.688 \\ 165.52 \end{pmatrix} \end{pmatrix}; \quad \bar{x}^* - \bar{m} = \begin{pmatrix} 3.104 \\ -40 \\ \begin{pmatrix} -9.312 \\ 65.52 \end{pmatrix} \end{pmatrix}; \quad \bar{\sigma} = \begin{pmatrix} 10 \\ 13 \\ \begin{pmatrix} 300 \\ 30 \end{pmatrix} \end{pmatrix}; \quad \bar{\lambda} = \begin{pmatrix} 5{,}000 \\ 10{,}000 \\ 200{,}000 \end{pmatrix}$$

**(38)**

$$\bar{b} - A\bar{x}^* = \begin{pmatrix} 0 \\ 0 \\ 72.312 \end{pmatrix}$$

Таким образом, оптимальный план состоит в:

- По первому независимому продукту перевыполнении плана в 30 кг. на 3.104 кг., при допуске ± 10 кг. и точной стоимости 5,000 руб. за 30 кг. первого продукта.
- Отказе от производства и поставки второго независимого продукта.
- По комплекту из третьего и четвёртого продуктам: недовыполнение плана в 900 кг. на 9.312 кг. при допуске ± 300 кг и перевыполнении плана в 100 кг. на 65.52 кг., при допуске ± 30 кг. и общей точной стоимости 200,000 руб. за комплект (900 кг,100 кг) третьего и четвёртого продуктов.
- Полная Стоимость найденного оптимального плана $F(\overline{x}^*) = 197{,}812.64$ руб., что достаточно близко к «точной» стоимости плана: $F(\overline{m}) = 215{,}000$ руб.,
- Первый и второй ресурсы израсходованы полностью. Третий ресурс – избыток в 72.312 кг.

Проверим выполнение условий оптимальности для найденного решения $(\overline{x}^*, \overline{y}^*)$, т.е. условий Куна-Такера (8):

$$\overline{\nabla}_{\overline{x}} L(\overline{x}, \overline{y}) = \overline{g}(\overline{x}) - \overline{y} \cdot A = \begin{pmatrix} 380.18 \\ 5.4 \\ 523.74 \\ 238.21 \end{pmatrix} - \begin{pmatrix} 380.18 \\ 317.61 \\ 523.74 \\ 238.21 \end{pmatrix} = \begin{pmatrix} 0 \\ -312.21 \\ 0 \\ 0 \end{pmatrix} \le \overline{0}$$

$$\overline{\nabla}_{\overline{y}} L(\overline{x}, \overline{y}) = \overline{b} - A \cdot \overline{x} = \begin{pmatrix} 49{,}500 \\ 9{,}900 \\ 5700 \end{pmatrix} - \begin{pmatrix} 49{,}500 \\ 9{,}900 \\ 5{,}627.688 \end{pmatrix} = \begin{pmatrix} 0 \\ 0 \\ 72.312 \end{pmatrix} \ge \overline{0}$$

**(39)**

$$(\overline{g}(x) - \overline{y} \cdot A)\overline{x} = \begin{pmatrix} 0 & -312.21 & 0 & 0 \end{pmatrix} \cdot \begin{pmatrix} 33.104 \\ 0 \\ 890.688 \\ 165.52 \end{pmatrix} = \overline{0}$$

$$\overline{y}(\overline{b} - A\overline{x}) = \begin{pmatrix} 7.94 & 12.67 & 0 \end{pmatrix} \cdot \begin{pmatrix} 0 \\ 0 \\ 72.312 \end{pmatrix} = \overline{0}$$

$$\bar{x}^* = \begin{pmatrix} 33.104 \\ 0 \\ 890.688 \\ 165.52 \end{pmatrix} \geq 0 \qquad \bar{y}^* = \begin{pmatrix} 7.94 \\ 12.67 \\ 0 \end{pmatrix} \geq 0$$

Таким образом, в найденной точке $(\bar{x}^*, \bar{y}^*)$, все соотношения Куна-Такера выполняются.

Выпишем балансовые соотношения (13) – (15):

**(13`)** $G(\bar{x}^*) = \bar{g}(\bar{x}^*) \cdot \bar{x}^* = 518{,}502.33 \approx 518{,}502.36 = \bar{y}^* \cdot (A \cdot \bar{x}^*)$

Т.е., внутренняя стоимость ресурсов равна внешней градиентной (маргинальной), стоимости $G(\bar{x}^*) = \nabla_{\bar{x}} F(\bar{x}^*) \cdot \bar{x}^*$ продукции в точке $\bar{x}^*$. (Ошибка в 3 копейки – обусловлена несовершенством алгоритма решения задач (17) и (20))

**(14`)** $\bar{y}^* \cdot (A \cdot \bar{x}^*) = 518{,}502.36 = \bar{y}^* \cdot \bar{b}$

- Полная внутренняя стоимость израсходованных ресурсов, равна полной внутренней стоимости, начального запаса ресурсов $\bar{b}$.

**(15`)** $G(\bar{x}^*) = \bar{g}(\bar{x}^*) \cdot \bar{x}^* = 518{,}502.33 \approx 518{,}502.36 = \bar{y}^* \cdot \bar{b} = \bar{y}^* \cdot (A \cdot \bar{x}^*)$

- Полная внешняя градиентная стоимость произведённой продукции $G(\bar{x}^*) = \bar{g}(\bar{x}^*) \cdot \bar{x}^*$ равна полной внутренней стоимости $\bar{y}^* \cdot \bar{b}$ (в двойственных линейных ценах $\bar{y}^*$) начального запаса ресурсов $\bar{b}$, или израсходованных ресурсов $A \cdot \bar{x}^*$.

Как отмечено в **Замечании 8**, полная внутренняя стоимость израсходованных $\bar{y}^* \cdot (A \cdot \bar{x}^*)$ (или начальных $\bar{y}^* \cdot \bar{b}$) ресурсов равна внешней градиентной (маргинальной), стоимости $G(\bar{x}^*)$ продукции в точке $\bar{x}^*$, а не внешней полной стоимости $F(\bar{x}^*)$ плана $\bar{x}^*$:

$$\bar{y}^* \cdot \bar{b} = \bar{y}^* \cdot (A \cdot \bar{x}^*) = 518{,}502.36 \approx 518{,}502.33 = G(\bar{x}^*) = \bar{g}(\bar{x}^*) \cdot \bar{x}^* \neq 197{,}812.64 = F(\bar{x}^*)$$

Приведённый пример показывает, как значительно может отличаться внутренняя градиентная стоимость ресурсов от полной интегральной внешней стоимости плана.

**Замечание 13.** Критерий $F(\bar{x}^*)$ является оценочной структурой, отражающей субъективную оценку полезности продукции с точки зрения внешнего, по отношению к данной экономической системы, субъекта экономики - верхнего уровня управления, или другого экономического субъекта. Полученные двойственные оценки $\bar{y}*$ – являются внутренними субъективными линейными градиентными (маргинальными) ценами на ресурсы.

Вопросы построения субъективной нелинейной функции распределения стоимости и функции плотности распределения стоимости на ресурсы, как реакции экономической подсистемы на внешние требования $F(\bar{x})$, будут рассмотрены в следующей статье.

В заключение автор выражает большую благодарность А.М.Тульской за полезные обсуждения и поддержку в самом начале работы в этом направлении, а так же М.С.Недьяк, прочитавшему статью в рукописи и сделавшему ряд ценных замечаний и советов.

**Библиография:**

1. Математическое Программирование. В.Г.Карманов, Москва, «Наука», 1986.
2. Теория Вероятностей. Е.С.Вентцель. Москва, «Наука», Физматгиз, 1969.